# Developing AMS measurement of $^{59}$Ni at CIAE[*]

He Ming(何明)[1], Wang Wei(王伟)[2], Ruan Xiangdong(阮向东)[3], Li Chaoli(李朝历)[1], Dong Kejun(董克君)[1], Du liang(窦亮)[1], Xie Lin Bo(谢林波)[1], Li Zhenyu(李振宇)[1], Zhen Guowen(郑国文)[1], Hu Hao(胡豪)[1], Liu Jiancheng(刘建成)[1], Jiang Shan(姜山)[1]

[1] China Institute of Atomic Energy, P.O.Box 275(50), Beijing 102413, China

[2] China National Nuclear Corporation，Beijing 100822,China

[3] College of Physics Science and Technology, Guangxi University, Nanning 530004, China

**Abstract:** Accelerator mass spectrometry(AMS) measurement of $^{59}$Ni has been established at CIAE with the HI-13 Tandem Accelerator and the recently developed ΔE-Q3D detection system. $^{59}$Ni standard and commercial NiO samples were measured to check the performance of the ΔE-Q3D detection system on $^{59}$Ni isobar separation and suppression. An overall suppression factor of about $10^7$ for the interfering isobar $^{59}$Co resulting in detection sensitivity as low as $3.8\times10^{-13}$ atomic ratio ($^{59}$Ni/Ni) has been obtained. Based on these techniques, the AMS measurement method of $^{59}$Ni with high sensitivity is developed.

## 1  Introduction

The long lived radioisotopes of $^{59}$Ni with a half life of 76ka can be used in a number of applications including low-level radioactive waste management[1], cosmic radiation study[2], neutron dosimetry[3,4] and astrophysics[5,6]. Due to the long half life and pure electron capture decay and no gamma emitting, it is very difficult to measure $^{59}$Ni with decay counting method. Accelerator mass spectrometry(AMS) is the only technique to measure $^{59}$Ni at low concentration. The main problem in AMS

---

[*] supported by National Natural Science Foundation of China (11175266 and 10875176)

1)E-mail:minghe@ciae.ac.cn

measurement of $^{59}$Ni is the interference of the stable isobar $^{59}$Co, whose concentration in the purified samples is difficult to be reduced to less than $10^{-6}$g/g by chemical method. So, in order to get high sensitivity of measurement of $^{59}$Ni, the technique for removing the interference of $^{59}$Co should be developed in the AMS system. Several methods have been developed in other AMS labs [6, 7]. The LLNL AMS Group developed the projectile X ray method for identifying $^{59}$Co and $^{59}$Ni according to their different K$\alpha$ X ray energies [7]. However, due to the limited identification power and detection efficiency, the sensitivity is $\sim 10^{-11}$ atomic ratio ($^{59}$Ni/Ni). A gas-filled magnet (GFM) combined with multi-anodes gas ionization chamber has been developed for $^{59}$Ni AMS measurement in Munich AMS Group based on their 14MV MP tandem accelerator [8]. In that method, the $^{59}$Ni and $^{59}$Co were separated at the focal plane of the GFM due to their different mean charge states in the GFM, and then gas ionization chamber was used to further identify $^{59}$Co and $^{59}$Ni. Based on this technique the sensitivity of $\sim 10^{-14}$ atomic ratio ($^{59}$Ni/Ni) was achieved. Recently, a $\Delta$E-Q3D detection system has been developed on HI-13 tandem accelerator for the measurement of medium mass heavy nuclides at China Institute of Atomic Energy (CIAE) [8]. The details of individual optical elements in the AMS setup can be found in Ref. [9]. The $\Delta$E-Q3D detection system mainly consists of a Q3D magnetic spectrometer with an absorber at its entrance and a multi-anode gas-ionization chamber in its focal plane. The Q3D magnetic spectrometer has the advantages of high energy resolution of $1.8\times 10^{-4}$, large dispersion of 11.3cm(1% $\Delta$P/P with P being the momentum), large spatial angle and large kinematics compensating ability[10]. Based on this system, $^{32}$Si and $^{53}$Mn were successfully measured at CIAE[9,11]. The principle of this method is described in Ref. [8]. Briefly speaking, a very homogeneous membrane with suitable thickness is used as an absorber and placed at the entrance of the Q3D magnetic spectrometer. Isobars lose different amounts of energy (ΔE) after passing through the absorber. Then the isobars are separated at the focal plane of the Q3D magnet spectrometer due to their different residual energies. The nuclide of interest, having the charge state with the highest stripping probability after passing through the absorber, is selected and focused onto a designated position of the focal plane and recorded by a multi-anode ionization chamber, while the interfering isobars are largely rejected. Tails of the isobars and other interferences, due to the straggling and scattering effects, are further rejected with the ionization chamber. The combination of the ΔE-Q3D and multi-anode ionization chamber is proved to be very effective for eliminating the $^{59}$Co isobaric interference problem on the measurement of $^{59}$Ni.

## 2. AMS Measurement of $^{59}$Ni

### 2.1 Ion beam transport

The sample form of NiO was adopted for the AMS measurement of $^{59}$Ni at CIAE. The NiO powder was mixed with same weight of pure Ag powder (99.99%) for improving thermal and electric conductivity and pressed into Al target holders of a 40-sample NEC MC-SNICS ion source for AMS measurement. Typically, 1.5 μA of $^{58}$Ni$^-$ was extracted from the source. After passing through the electrostatic analyzer and injection magnet, $^{59}$Ni$^-$ (and $^{59}$Co$^-$) ions were selected for injection into the HI-13 Beijing Tandem Accelerator which was set at 11.5 MV. Carbon foil (3 μg·cm$^{-2}$) stripping was employed to produce positive atomic ions with high charge states and to break up other molecular ions. The positive ions were further accelerated by the same terminal voltage. A 90° double-focusing analyzing magnet was used to select $^{59}$Ni$^{12+}$ (and $^{59}$Co$^{12+}$) with energy of 149.5 MeV. None of these processes can separate $^{59}$Ni and the stable isobar $^{59}$Co. After switching magnet the isobar $^{59}$Co$^{12+}$ with the same energy as $^{59}$Ni$^{12+}$ were transported to the ΔE-Q3D detection system.

*2.2 Isobar separation*

A very homogeneous Si$_3$N$_4$ foil with thickness of 4.5 μm (four foils with thickness of 1 μm each and one foil with 0.5μm lapped over) was placed at the entrance of the Q3D magnet spectrometry as an absorber. After $^{59}$Ni and $^{59}$Co passing through the absorber, ions with the charge state of 19+, which offered a stripping probability of about 20%, were selected and transported to the focal plane of the Q3D magnet spectrometry. According to calculation based on TRIM[13], the energy loss of $^{59}$Ni and $^{59}$Co in the absorber are 45.1 MeV and 43.2MeV, respectively. The energies of $^{59}$Ni and $^{59}$Co after passing the absorber are 104.4MeV and 106.3MeV, respectively. This energy difference will cause 110mm separation between $^{59}$Ni and $^{59}$Co at the focal plane of the Q3D. A movable surface barrier detector (SBD) with a diameter of 12 mm was placed on the focal plane of Q3D to measure the distribution of $^{59}$Ni and $^{59}$Co. A sample with a $^{59}$Ni/Ni ratio of 2.0×10$^{-9}$ was used to check the separation between $^{59}$Ni and $^{59}$Co at the focal plane. Fig. 1 shows the position distribution of $^{59}$Co and $^{59}$Ni along the focal plane. Due to the strong background of $^{59}$Co and low count rate of $^{59}$Ni, no obvious $^{59}$Ni peak response is recognized by the SBD along the focal plane. $^{60}$Ni ions with the same magnetic rigidity as that for $^{59}$Ni ions were used for simulating the position of $^{59}$Ni. The result was shown in Fig.1.  Fig. 1 shows that the peaks of $^{59}$Ni and $^{59}$Co were separated by about 130mm on the focal plane which is little higher than the theoretical estimation, and each peck had a width about 60mm(FWHM). A four-anodes gas ionization chamber with an entrance widow of 100mm(width)×40mm(height) was mounted at the Q3D focal plane as shown in fig.1. By suitable adjusting the magnetic field of Q3D, about 95% of $^{59}$Ni can enter the gas ionization chamber while the $^{59}$Ni, which the peak position is about 90mm away from gas ionization chamber entrance window (fig.1), can be much reduced. By this way, most of the $^{59}$Co are eliminated. A suppression factor of more than 500 was achieved for $^{59}$Co ions.

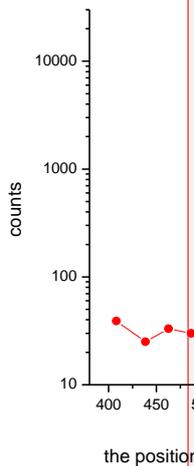

Fig.1 The position distribution of $^{59}$Co and $^{59}$Ni along the focal plane of Q3D magnetic spectrometer, $^{60}$Ni ions with the same magnetic rigidity as that for $^{59}$Ni ions were used for simulating the position of $^{59}$Ni. The position of the gas ionization chamber entrance window is also shown.

*2.3 Particle identification*

Although most of the $^{59}$Co was eliminated, small part of $^{59}$Co, can still enter the gas detector. According to the difference in energy losses of $^{59}$Co and $^{59}$Ni ions in the gas detector medium, $^{59}$Co and $^{59}$Ni can be identified with the four-anode gas ionization chamber further. Five signals, consisting of four signals ($\Delta E_1$, $\Delta E_2$, $\Delta E_3$, $E_4$) from the four anodes and one total energy signal ($E_t$) from the cathode, were used. A multi-parameter data acquisition system was used to identify $^{59}$Co and $^{59}$Ni with appropriate gates on these five signals. Fig.2 shows the plot of energy loss of $\Delta E_1$ verses $\Delta E_4$, $\Delta E_2$ verses $\Delta E_3$ for a laboratory standard with a $^{59}$Ni/Ni atomic ratio of 2.0×10$^{-9}$. The same spectra for a blank sample are shown in Fig.3. It can be seen from the standard sample spectra that $^{59}$Ni and $^{59}$Co are identified by the detector. More than 90% of $^{59}$Ni counts can be extracted with appropriate gates on five signals from the detector as shown in Fig. 2-c. At the same time, the background can be greatly removed. As shown in Fig. 3, more than 1.58×10$^5$ $^{59}$Co counts (Fig.3 -a) are accumulated with eight counts in the $^{59}$Ni peak region after applying all the same gates as in standard sample, corresponding to a $^{59}$Co suppression factor of 1.8×10$^4$ in the detector. An overall suppression factor of about 10$^7$ for $^{59}$Co can be obtained by using the ΔE-Q3D detection system.

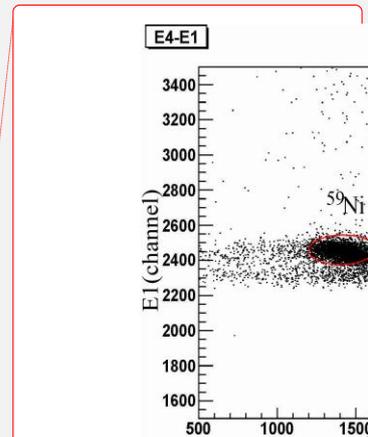

Fig.2-a.

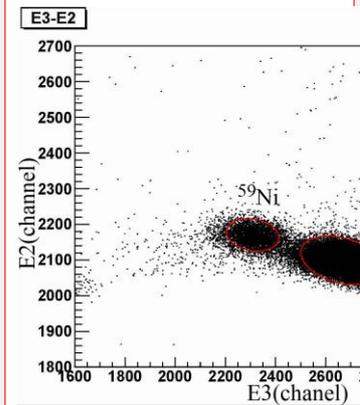

Fig.2-b.

Fig. 2. Two-dimensional spectra for a laboratory standard with $^{59}$Ni/Ni atomic ratio of 2.0×10$^{-9}$. Fig.2-a and Fig.2-b are the two-dimensional spectrum of $\Delta E_1$ versus $\Delta E_4$ and $\Delta E_2$ versus $\Delta E_3$ respectively, Fig.2-c is the two-dimensional spectrum of $\Delta E_1$ versus $\Delta E_4$ after the application of all the gates.

Fig.3. Two-dimensional spectra of $\Delta E_1$ versus $\Delta E_4$ (Fig.3-a) and $\Delta E_2$ versus $\Delta E_3$ (Fig.3-b) for a blank sample, Fig.3-c is the two-dimensional spectrum of $\Delta E_1$ versus $\Delta E_4$ after the application of all the gates.

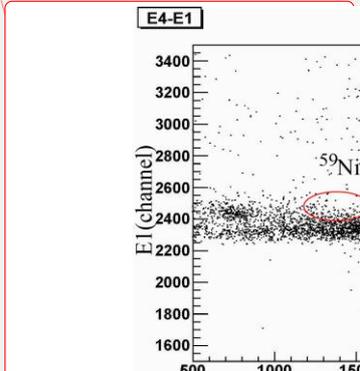

Fig.3-a.

## 2.4 Simultaneous measurement

The beam current of stable isotope Ni is another value which should be measured to get the atomic ratio of $^{59}$Ni/Ni. Usually, the beam current is unstable during the measurement of $^{59}$Ni. In order to avoid this uncertainty, a simultaneous measurement of $^{59}$Ni or $^{60}$Ni was developed. Fig.4 shows the schematic diagram of the injection system. After Ni$^-$ passes through the injection magnet, the $^{59}$Ni$^-$ was transported to the accelerator, meanwhile $^{58}$Ni$^-$ and $^{60}$Ni$^-$ were inward and outward deflected relative to

$^{59}$Ni$^-$, two offset Faraday cups at the image point of the magnet were used to record the beam current. In this way, $^{60}$Ni(or $^{58}$Ni) and $^{59}$Ni were measured simultaneously based on the $^{60}$Ni current measured in the offset Faraday cup and the counts of $^{59}$Ni recorded by the detector after software gate coincidence and background subtraction. The standard samples and real samples can be measured in turn for normalizing the transmission and detection efficiency. After normalizing by the standard sample values, the results of real samples can be obtained.

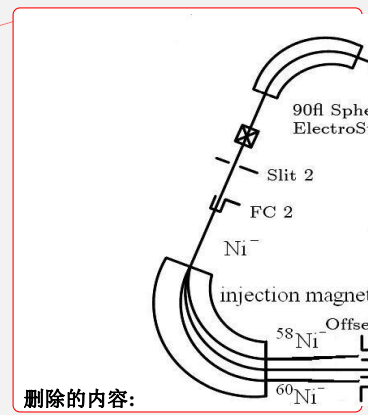

Fig.4. The schematic diagram of injection system.

## 3 Result and discussion

Two laboratory standard samples and one blank sample were measured with the established method. The $^{59}$Ni laboratory standard samples were produced via irradiation of natural Ni with thermal neutrons. Two laboratory standard samples with $^{59}$Ni/Ni nominal atomic ratio of $(2.02\pm0.17)\times10^{-9}$ (Ni-S-1) and $(2.51\pm0.22)\times10^{-11}$ (Ni-S-2) which were produced by chemical dilution from the irradiated sample were measured. A commercial NiO blank sample was also measured to check the background level. The results are presented in Table 1. The atomic ratios of $^{59}$Ni/Ni were normalized to the sample of Ni-S-1 with the value of $2.02\times10^{-9}$. The uncertainties of the normalized ratio come from the AMS systematic error, counting statistics, background of $^{59}$Co and beam current measurement. The background level of $3.8\times10^{-13}$ atomic ratio ($^{59}$Ni/Ni) was obtained from the blank sample.

Table1. Results of $^{59}$Ni/Ni ratios for laboratory standards and a commercial blank, normalized to the sample of Ni-9 with the value of $2.02\times10^{-9}$.

| Sample | Nominal Ratio ($^{59}$Ni/Ni) | Measurement Time(s) | $^{59}$Ni Counts | $^{60}$Ni$^-$ Mean Current (nA) | Normalized Ratio ($^{59}$Ni/Ni) |
|---|---|---|---|---|---|
| Ni-S-1 | $2.02\times10^{-9}$ | 60 | 2889 | 429.3 | $(2.02\pm0.12)\times10^{-9}$ |
| Ni-S-2 | $2.51\times10^{-11}$ | 600 | 283 | 313.5 | $(2.74\pm0.21)\times10^{-11}$ |
| Blank | \ | 1000 | 8 | 388.6 | $(3.8\pm1.3)\times10^{-13}$ |

A measurement method of $^{59}$Ni in the CIAE-AMS facility was developed. The combination of ΔE-Q3D and multi-anode gas ionization chamber is an effective way to remove the $^{59}$Co background. Further reduction of $^{59}$Co interference is the key to further improve the AMS measurement sensitivity of $^{59}$Ni. In the next step, the following aspects will be developed for depressing $^{59}$Co interference: (1) More comprehensive chemical procedures will be studied for removing Co content in the sample material.(2) Reducing the size of the entrance window of gas ionization chamber to $75\times40$ mm$^2$ to further depress $^{59}$Co background while still recording nearly 80% of $^{59}$Ni events. Based on the above improvements, the detection sensitivity of $10^{-14}$ for $^{59}$Ni/Ni atomic ratio is expected to be achieved with the CIAE-AMS system. Applications based on the $^{59}$Ni measurement will be carried out in the next step.